\documentclass[a4paper, 18pt]{article}
\usepackage[margin=0.75in]{geometry}
\usepackage{amsmath}
\usepackage{graphicx}
\usepackage{natbib}
\usepackage{hyperref}
\usepackage{cleveref}

\begin{document}
\noindent {\LARGE \textbf{Axonal Computations}}
\vspace{14pt}

\noindent {\Large Pepe Alcami$^{1,2,*}$, Ahmed El Hady$^{3,4,*}$} \\

\noindent{\small $1$. Neurobiology, Ludwig-Maximilians University, Martinsried, Munich, Germany

\noindent $2$. Behavioral Neurobiology, Max Planck Institute for Ornithology, Seewiesen, Germany

\noindent $3$. Princeton Neuroscience Institute, Princeton University, Princeton , New Jersey , USA

\noindent $4$. Howard Hughes Medical Institute, Princeton University, Princeton , New Jersey , USA

\noindent *correspondence should be addressed to Pepe Alcami (alcami@bio.lmu.de)or  Ahmed El Hady (ahady@princeton.edu)}
\vspace{14pt} \newline

\noindent {\Large \textbf{Abstract}}\newline

Axons functionally link the somato-dendritic compartment to synaptic terminals. Structurally and functionally diverse, they accomplish a central role in determining the delays and reliability with which neuronal ensembles communicate. By combining their active and passive biophysical properties, they ensure a plethora of physiological computations. In this review, we revisit the biophysics of generation and propagation of electrical signals in the axon, their complex interplay, and their rich dynamics. We further place the computational abilities of axons in the context of intracellular and intercellular coupling. We discuss how, by means of sophisticated biophysical mechanisms, axons expand the repertoire of axonal computation, and thereby, of neural computation.

\section{Introduction}
Neurons are compartmentalized into input compartments formed by dendrites and somas, and an output compartment, the axon. However, in the current era, there is a widespread tendency to consider that the biophysics of single neurons do not matter to understand neuronal dynamics and behavior. Neurons are often treated as point processes with disregard of the complex biophysical machinery that they have evolved. Moreover, neuronal computations are assumed to be mostly performed by dendrites or at synapses  \citep{SUDHOF2008469, stuart2016dendrites}, and axons are reduced to simple, static and reliable devices. However, a wealth of literature supports that this is not the case: axons form complex structures that ensure a variety of sophisticated functions. Here, we aim to review evidence that axons perform complex computations which depend on a myriad of biophysical details and ensure the generation and propagation of neuronal outputs. We will not discuss biophysics of synaptic release, reviewed elsewhere \citep{SUDHOF2008469}.

More than six decades after seminal discoveries in experimentally accessible invertebrate axons \citep{Hodgkin1952}, axonal research has unraveled previously-unsuspected, rich and dynamical electrical signaling in axons, which consists of a hybrid of analog and digital signaling. Contrary to the giant invertebrate axons studied in the early days (Fig. 1A; \citep{Hodgkin1952, Furshpan1959}), most invertebrate and vertebrate axons are thin and  present complex extended arborizations (e.g., Fig. 1B-D), making it difficult to record from them. Electron and optical microscopy have made it possible to deepen our understanding of fine axonal structures \citep{Rash2016, D'este2017}. Moreover, the combination of new structural imaging techniques with electrophysiology has made it possible to study how structural changes at the sub-micrometer scale impact function \citep{chereau2017superresolution}.

We will first illustrate general signaling principles in axons, before delving into the generation and propagation of action potentials (APs) and their dynamical regulation. Finally, we will position axons in the context of their interactions with other compartments and with each other, by virtue of axo-axonal coupling \citep{Katz1940,Furshpan1959}  and finally, indirectly via glial cells.


\section{Review of Axonal Computations}


\subsection{GENERAL PRINCIPLES: SIGNALING IN AXONS}

\subsubsection{Of axons and brains
}
The propagation of electrical signals along axons controls the reliability and the timing at which neural networks communicate. The spatial extent of axonal trees introduces delays between the generation of an electrical signal (APs or subthreshold signals) in a neuron and the arrival of this signal to the presynaptic site, where information is relayed to a postsynaptic cell. Thereby, the axonal propagation delay influences the temporal relationship of presynaptic and postsynaptic activity \citep{Izhikevich2004}.

The delays in AP propagation encode information \citep{Seidl70} and contribute to re-configuring neural circuits through plastic mechanisms such as spike timing dependent plasticity \citep{bi1998synaptic, Izhikevich2004}. Each synapse has its own 'critical window', given by the delay between presynaptic and postsynaptic activity, to induce synaptic plasticity. The speed at which signals travel along axons can vary over four orders of magnitude, from tens of centimeters per second in thin unmyelinated axons to hundreds of meters per second in giant myelinated fibers  \citep{Schmidt-hieber2008,Xu1979}); highlighting the ability of biophysical specializations characterizing different axons to conduct APs at different speeds.

Axons have evolved different intricate geometries (Fig.1), revealing specific structure-function specializations. Constraints to axonal morphology include (1) spatial constraints: axons occupy large volumes of nervous system, yet they require internal components to their function such as mitochondria, which constrain their minimal functional size; (2) energetic requirements may have evolutionary exerted a selection pressure on axonal properties \citep{Perge7917, Harris356}; (3)  efficiency of electrical information processing in relation to  structure-function specializations. The last point on electrical information processing will be the main focus of the current review.

\subsubsection{Propagation of electrical signals in the axon}
Propagation of electrical signals in the axon results from a combination of specialized active and passive mechanisms. Active properties are shaped by neuronal voltage-gated ion channels recruited as a function of the dynamics of membrane potential whereas passive properties are determined by the axonal membrane at rest and by axonal geometry.

As increasingly appreciated, APs are not just an on/off switch, and electrical signaling in the axon should be regarded as a hybridization of analog and digital signalling \citep{Shu2006}. Furthermore, APs are strongly regulated by the background analog activity provided by subthreshold postsynaptic events. For example, the inactivation of potassium $K_{v}$1 channels in the axon initial segment broadens the axonal AP waveform and increases unitary excitatory postsynaptic potentials (EPSP) amplitude in layer V pyramidal neurons (Kole et al. 2007). Note that some neurons do not encode information with APs but only with graded signals (graded potential neurons)\citep{borst1996intrinsic} .

\subsubsection{Non-electrical signaling}
Signaling in axons and neurons is usually regarded as purely electrical. However, electrical signals are  accompanied by other biophysical changes in the neuronal membrane. The AP is accompanied by changes in many biophysical properties such as temperature, mechanical membrane properties and optical birefringence \citep{cohen1970changes, tasaki1992heat, howarth1975heat,howarth1975heat2, abbott1965initial, tasaki1982rapid}. For example the changes in the optical properties of the axon during propagation have been at the basis of the label free interferometric imaging of APs in in-vitro systems \citep{oh2012label, akkin2007depth, batabyal2017label}.

Mechanical displacements associated with the AP have been measured in many experimental contexts \citep{tasaki1982further, tasaki1982rapid, iwasa1980mechanical,hill1977laser}. These mechanical displacements can be regarded as propagating surface modes that are elicited via the large electrostatic force produced by the AP. In this regard, the AP is an electro-mechanical pulse \citep{el2015mechanical}. Converging evidence suggests that mechanics play a role in electrical signalling \citep{tyler2012mechanobiology}. Moreover, there is mounting evidence that some voltage sensitive channels such as sodium and potassium channels are mechanically modulated locally  and that many neurons express mechanically-activated channels \citep{ranade2015mechanically, schmidt2008voltage, schmidt2012mechanistic}. In addition, mechanical pressure at the soma of high-resistance neurons can induce APs \citep{alcami2012measuring}, an effect likely to be even stronger at axons, which are highly resistive. Theoretical and biophysical evidence of non-electrical signaling has surprisingly not yet led to a thorough investigation of the functional relevance of these mechanical displacements and other non-electrical signaling modalities to axonal computation.

\subsection{BIOPHYSICS OF ACTION POTENTIAL GENERATION}
\subsubsection{A brief historical perspective on the action potential}
At the resting, non-excited state, mostly potassium channels are open and the resting potential is, as a consequence, close to the reversal potential for potassium, maintained around -70 mV. In 1939, Hodgkin and Huxley published the first trace of an AP recorded from the squid axon using an intracellular electrode where one can see a very clear overshoot. Following this in 1949, the proposal that sodium ions are the main mediator of AP generation was put forward by Hodgkin and Katz \citep{hodgkin1949effect}. They studied the effect of systematically varying the concentration of sodium ions and measured its impact on the amplitude of the AP recorded. Subsequently, Keynes managed to show that nerve excitation leads to an increase in the transmembrane flow of sodium ions by tracing the movement of the radioactive isotope $Na^{24}$ in repeatedly stimulated squid axons \citep{keynes1951ionic}. These seminal findings confirmed that sodium ions are the main contributors to AP generation. 

\subsubsection{Variations of action potentials}

APs are characterized by a width, height and an overshoot magnitude (i.e. referring to the AP height beyond 0 mV). Despite sharing a general biophysical mechanism for initiation, the shape of APs varies in different types of neurons. For example, there are cells that exhibit very narrow APs of a few hundreds of microseconds in width, reflecting a fast spiking behavior, such as some GABAergic interneurons, Purkinje neurons which are GABAergic projection neurons, glutamatergic neurons of the subthalamic nucleus and Medial nucleus of the trapezoid body (MNTB) cells in the superior olivary complex. CA1 pyramidal neurons in the hippocampus have, in contrast, relatively wide AP and dopaminergic neurons have an even wider AP of up to 4 ms ( reviewed in \citep{bean2007action}). 
Apart from the shape of the single spike, neurons can exhibit a diversity of firing patterns. For example, they can be bursting or non-bursting on one hand and they can be adapting or non-adapting on the other hand. Adaptation during a train of APs refers to the process by which AP properties, typically rate and amplitude, decrease within the spike train. There are many biophysical mechanisms by which spike frequency adaptation can happen. The most prominent mechanisms are an increase in outward current flowing through calcium-activated potassium channels and an increasing outward current produced by the electrogenic sodium-potassium pump \citep{powers1999multiple}. Moreover, there is also a substantial sodium channel inactivation induced by a long lasting depolarization \citep{sawczuk1997contribution}. The diversity of AP shapes along with the firing patterns are a result of the different combinations of ion channels that the neuron expresses (reviewed in \citep{marder2006variability}).

\subsubsection{The axon initial segment}

Before delving into the mechanistic details of AP generation, it is crucial to appreciate the complex anatomy of the Axon Initial Segment (AIS), the site of AP initiation (Fig. 3A). Generally speaking, the AIS refers to the first 20-60 $\mu$m of the axon but the specific length of the AIS is variable across neurons as it has to adapt it to its own excitability properties. The excitability of the AIS is modulated by varying compositions and subtypes of sodium and potassium channels clustered at the AIS.  As an example, motor alpha and gamma neurons have an AIS length of 28 - 29 $\mu$m but alpha and gamma motor neurons differ in the percentage of $Na_{v}$ 1.6 and $Na_{v}$ 1.1 channels subtypes distribution  \citep{duflocq2011characterization}   which underlies the different excitability properties of both subtypes of motor neurons. It is important to note that the different excitable properties are owed to the fact that $Na_{v}$ 1.6 and $Na_{v}$ 1.1 channels have different kinetics. 

The AIS contains a highly-specialized protein machinery that gives it a distinct character. One such protein is Ankyrin G, a scaffolding protein that is also present at the nodes of Ranvier \citep{kordeli1995ankyrin}. Ankyrin G anchors voltage-gated channels such as voltage-gated sodium ($Na_{v}$) and potassium channels ($K_{v}$) to the membrane along with other adhesion molecules \citep{davis1996molecular}. Beta IV spectrin is another protein expressed in the axon initial segment. Beta IV spectrin's main function is to cluster sodium channels at the axon initial segment while simultaneously binding to the actin cytoskeleton (reviewed in \citep{rasband2010axon})( Fig.3 A).

The AIS in mammalian neurons has been established as the site of the AP initiation following a series of seminal studies that began in the mid 1950s \citep{araki1955response, coombs1957generation, fatt1957electric}. The location of AP initiation has been further confirmed by combining precise electrophysiological measurements and imaging technologies. These allow to precisely identify the locus of AP initiation which was established to be in the distal part of the AIS, 20 to 40 $\mu$m from the soma \citep{atherton2008autonomous, foust2010action, kole2007axon, Meeks2007, Palmer2010, Palmer1854}. The AIS is considered to be the site of AP initiation also because of several key properties. It contains a high sodium channel density. Furthermore, AIS sodium channels show a voltage dependence shifted to lower voltages, favoring their activation at less depolarized voltages than at the soma. Finally, the relatively large electrotonic distance of the AIS from the capacitive load of the soma renders distal sodium influx more efficient in evoking a local membrane depolarization, compared to an AIS that would start at the soma. Note that electrophysiological recordings from invertebrates have shown that the initiation of the AP can happen at multiple locations acting in an independent manner \citep{CALABRESE1974316, MARATOU20001, Meyrand2803}. Along with its specialized protein machinery and acting as the site of AP initiation, the AIS also acts as a diffusion barrier between the somatodendritic and axonal compartments, filtering transport materials passing from soma to axon \citep{song2009selective, brachet2010ankyrin}. Note however that the AIS length can be regulated in an activity-dependent manner \citep{Kuba3443} and that in some neurons, AP initiation has been reported to occur at the first node of Ranvier \citep{Lehnert5370, Clark2005} , as will be developped later.

\subsubsection{The sodium ionic dynamics and action potential initiation}
The extent to which the density of sodium channels is higher in the AIS and the contribution of this high density of sodium channels to AP initiation is a matter of active investigation. There is a consensus that sodium channel density is higher in the AIS but the order of magnitude is still unclear. Immunostaining of sodium channels consistently indicates that there is higher channel density in the AIS of various neuron types \citep{Meeks2007, Wollner8424, Boiko2306}. Lorincz and Nusser in \citep{Lorincz906} counted around 200 $Na_{v}$ 1.6 sodium channels per micrometers square in the AIS of hippocampal pyramidal cells using electron microscopy. Given that the conductance of a single $Na_{v}$ channel is around 15 pS ( Colbert and Johnston, 1996), one would expect a conductance density of about 3000 pS per micrometers square. On the contrary, electrophysiological measurements from membrane patches pinpoint that sodium channel density is about 3 – 4 channels per micrometers square in the AIS, which is the same as the somatic density (Colbert and Johnston 1996, Colbert and Pan 2002). A similar conclusion of equal densities was reached on the basis of recordings from blebs that form when cortical axons are cut. This might be due to the inability to draw AIS $Na^{+}$ channels into the patch-clamp recording pipette due to their tight coupling to the actin cytoskeleton \citep{Kole2008}. In this article, the authors record a much larger sodium current after disruption of the actin cytoskeleton \citep{Kole2008}.

Kinetics of sodium currents underlying AP generation have been extensively studied. Hodgkin and Huxley have proposed that the activity of sodium channels can be fitted by $m^{3}$ activation kinetics. Baranauskas and Martina (2006) \citep{Baranauskas671} found that sodium currents in three types of central neurons (prefrontal cortical cells, dentate gyrus granule cells and CA1 pyramidal cells) activate faster than predicted by Hodgkin-Huxley type kinetics following $m^{2}$ activation kinetics. Moreover, it was found that the half activation voltage of voltage-gated sodium channels in layer 5 pyramidal neurons is 7 -14 mV lower in the distal AIS compared to the soma and decreases further with increasing distance from the soma \citep{Colbert, Hu2009}. $Na_{v}$1.6 channels, which are predominantly expressed in the axon initial segment and have a lower half activation voltage \citep{Rush2005}, are proposed to be primarily responsible for the initial slope of the AP. Sodium channels at the AIS are more capable of producing a persistent sodium current \citep{Astman3465, STUART19951065}. The persistent sodium current has a significant influence on the AP threshold \citep{Kole2008}. Moreover, it is implicated in the generation of the AP afterdepolarization and it therefore contributes directly to the generation of high frequency AP bursts \citep{Azouz1996}.  

Using sodium imaging to follow sodium influx in axon, soma and basal dendrites \citep{fleidervish2010na+}, the authors suggest that the ratios of $Na_{v}$ channel densities in these regions are approximately 3:1:0.3. Interestingly in another study by \citep{lazarov2018axon},  APs were initiated in the AIS, even when axonal $Na_{v}$ channel density was reduced to about 10 percent in a beta IV spectrin mutant mouse. This experimental finding indicates to a great extent that AP initiation in the AIS does not require such a high local channel density. However in that study, the precision of AP timing was substantially compromised when axonal channel density was reduced. Likewise, Kopp-Scheinpflug and Tempel found a decrease in the temporal accuracy of AP generation from MNTB cells in a beta IV spectrin mutant mouse \citep{Koppscheinpflug2015213} 

The aforementioned section highlights the complexity of sodium ion dynamics and begs for novel imaging modalities that allow tying ultrasfast dynamics with the axonal ultrastructures in order to get insights into the intricate biophysical mechanisms underlying the very first microseconds of AP initiation. 

\subsubsection{The action potential rapidness}
A simple but very informative way to study the properties of the APs is to plot the time derivative of the voltage (dV/dt) versus the voltage. This is called “phase-plane plot”. The spike threshold can be easily visualized in such a representation, where it corresponds to the voltage at which dV/dt rises abruptly. Coombs et al. \citep{coombs1957generation}  noticed that the main spike is preceded by a smaller earlier component (referred to as a “kink”). This component is interpreted as reflecting initiation of the spike in the initial segment of the axon. One can record such a component in somatic spikes in many central neurons, including neocortical pyramidal neurons which we will focus our discussion on here. 
As mentioned above, one of the most striking features of the AP recorded from cortical neurons has been the existence of a “kink” at the initiation of the AP (Fig. 3B1). One can define the rapidness of AP onset in this cell as the slope of the phase-plane plot at dV/dt = 10 mV /ms. In \citep{naundorf2006unique},  the AP rapidness was measured from the cat visual cortical cells. AP rapidness varied between around 20 to 60 $ms^{-1}$. It is important to mention that sharp, step-like onsets of APs have been recorded in vivo in many preparations (cat visual cortex: \citep{ Azouz2209}, and cat somatosensory cortex: in \citep{Yamamoto1990}). 

Several hypothesis have been proposed to explain the origin of the AP kink: 1) the backpropagation to the soma of a smoother AP generated at the AIS; 2) an abrupt opening of sodium channels due to the biophysics of neuronal compartmentalization; 3) a decreased membrane time constant due to the loading of the dendritic compartment and 4) the cooperativity of sodium channels at the AIS.

The “lateral current hypothesis” states that the “kink” at spike onset reflects lateral current coming from the axon which becomes sharper through backpropagation from the initiation site to the soma while initiation is smooth at the initiation site  \citep{Yu7260, McCormick2007} . In Yu et al. 2008, authors perform simultaneous recordings from axon blebs and soma, finding a smoother AP onset in the axon, additionally reproducing these results in a model. It is important to note that this study supporting the lateral current hypothesis was done in blebs recordings which are injured axons that may have undergone severe cytoskeletal reorganization  \citep{Spira2003} affecting sodium channels dynamics. This reorganization might alter the true dynamics of AP initiation. 

Although the kink indeed might reflect the lateral current coming from the axon \citep{Milescu12113}, this hypothesis fails to account for the ability of neurons to follow 200 - 300 Hz frequency inputs. In order to account for this discrepancy, Romain brette in \citep{Brette2013} proposed that the compartmentalization and the distance between the soma and the AIS leads to spike initiation sharpness. In his proposal, Brette suggests that the rapidness arises from the geometrical discontinuity between the soma and the AIS, rather than from the backpropagation of axonal APs. When sodium channels are placed in a thin axon, they open abruptly rather than gradually as a function of somatic voltage, as an all-or-none phenomenon.

Another proposal that takes into account the geometry of neurons is  Eyal et al. \citep{Eyal8063} in which the authors propose that increasing the dendritic membrane surface area (the dendritic impedance load) both enhances the AP onset in the axon and also shifts the cutoff frequency of the modulated membrane potential to higher frequencies. This “dendritic size effect” is the consequence of the decrease in the effective time constants of the neuron with increasing dendritic impedance load. The authors have shown this in a computational model of reconstructed layer 2 / 3 pyramidal neurons of humans and rats. The firing pattern at the axon is strongly shaped by the size of the dendritic tree. Authors predict that neurons with larger dendritic trees have a faster AP onset.

Another proposal to interpret AP onset rapidness is that  sodium channels, which are assumed to be opening independently within the Hodgkin-Huxley framework, are gated cooperatively. The cooperativity model proposed that the half-activation voltage of the channels becomes dependent on the probability of the opening of the neighboring channels. These cooperative effects might happen mechanistically on very fast timescales either through a purely electrical, mechanical or electro-mechanical coupling. Though there is no direct experimental test of the cooperativity of neuronal sodium channels at the AIS, it can theoretically account for the observed discrepancy between the sodium channel density and the very rapid rise of the AP at the site of initiation in the initial segment \citep{naundorf2006unique}. It is important to note that cooperative gating has been previously observed in calcium, potassium and HCN channels \citep{marx2001coupled, Kim20141661, dekker2006cooperative}.

Apart from its mechanistic underpinnings, the fast rise of APs has attracted both experimental and theoretical approaches to study its functional implications on the biophysics of neuronal populations. Before detailing those functional implications, it is important here to go through a useful theoretical abstraction: a typical cortical neuron, embedded in a cortical network in vivo, receives about 10.000 synaptic inputs. Assuming that each of these synaptic inputs is active with a rate on the order of 1 to 10 Hz, incoming signals arrive at a rate of 10 kHz. As a result, the membrane voltage exhibits strong, temporally irregular fluctuations. To understand the computational capabilities of e.g. cortical circuits, it is essential to characterize single neuron computation under such realistic operating conditions. To control the activity of entire neuronal circuits while preserving their natural firing characteristics, it would be advantageous to introduce artificial input components mimicking intrinsically generated synaptic input under precise experimental control. In order to study the dynamical properties of cortical neurons, experimenters have mimicked synaptic bombardment in vitro by injecting stochastic inputs modeled as an Ornstein-Uhlenbeck process in which sinusoidal inputs are embedded \citep{neef2013continuous, tchumatchenko2010correlations}. This experimental setting has allowed the measurement of the dynamic gain of neurons, which means how much neurons attenuate their input in the frequency domain and how fast they are able to follow a rapidly-fluctuating input. This has led to the establishment of the ability of cortical neurons to follow high frequency inputs up to 200-300 Hz \citep{higgs2009conditional, higgs2006diversity, Tchumatchenko12171}. Moreover, there has been a series of theoretical studies exploring the dependence of encoding capacity on the active properties of the AP initiation \citep{fourcaud2003spike, huang2012small, wei2011spike} .

\subsubsection{AP trajectories beyond the rapidness of initiation}

Although we have mostly concentrated initial spike rapidness, note that the phase plot can display a variety of trajectories after the onset of the AP (Fig.3). These trajectories are determined by additional ion channels in conjunction with sodium channels. Potassium channels are typically responsible for the repolarizing phase of the AP. The relative temporal profiles of activation of sodium and potassium channels and their subunit composition determine the width of the AP \citep{lien2003kv3}. Furthermore, the spatial location of ion channels also contributes to the shape of AP trajectories \citep{yang2016low}(Fig. 3B2). It has also been shown that the AP shape affects calcium currents and transmitter release in a calyx-type synapse in the rat auditory brainstem \citep{borst1999effect}. Therefore, the AP shape beyond the initial spike rapidness provides an extra dimension for information encoding on a variety of timescales.

\subsubsection{The action potential at nodes of Ranvier}

Although the AP is generated at the AIS, the nodes of Ranvier contain a machinery to regenerate the AP. It is important to note that there are striking similarities between axon initial segment and nodes of Ranvier. A great deal of the protein machinery in the axon initial segment is also present in the nodes of Ranvier where the regeneration of the AP is performed. Ankyrin G, the major scaffolding protein in the AIS and the ring-like arrangement of actin and beta IV spectrin are also found in the nodes of Ranvier \citep{d2017ultrastructural}. Nodes of Ranvier are distributed spatially along the axon to guarantee the faithful propagation of the AP. In addition, the first node of Ranvier was found to be crucial for high bandwidth bursting activity in neocortical layer 5 pyramidal neurons \citep{Kole2011}. In that study, authors show that nodal persistent sodium currents at the first node of Ranvier hyperpolarize AP threshold and amplify the afterdepolarization. The study opens up the space for a computational role of the first node of Ranvier beyond conduction of the propagating AP. The first node faithfully follows spike frequencies with a ∼100 $\mu$s delay \citep{foust2010action, Khaliq1935, Palmer2010, Palmer1854}.  This had led to the speculation that  the nodes of Ranvier, and in particular the first node, may have an active computational role in modulating the AP initiation itself.

It is worth noting that nodes of Ranvier also express $Na_{V}$ 1.6, the same sodium channel subtype that is expressed in the AIS. It is even more striking that the sodium channels at the nodes also undergo the developmental changes from $Na_{V}$ 1.2 to  $Na_{V}$ 1.6 during postnatal development, following a similar developmental trajectory as those found in the AIS \citep{Rios7001}. $Na_{V}$ is highly clustered at the nodes of Ranvier in the order of ~1200 channels per micrometers square \citep{Rosenbluth1976}, while internodes contain ~20–25 channels per micrometers square \citep{Ritchie1977}). This very high density ensures high-fidelity conduction of the AP which will be further developed  in the review. Given the close proximity of the first node to the cell body and high density of $Na_{V}$ channels, it has been postulated that, in addition to securing propagation, it could potentially generate the  AP \citep{Clark2005, Colbert, Lehnert5370} Although this might be happening, the overwhelming evidence favors that the AP initiation is happening at the AIS.

\subsubsection{Ectopic spiking}
AP initiation at the AIS and its subsequent orthodromic propagation have been extensively investigated. However, a number of studies demonstrates that distally generated spikes or ‘ectopic’ APs co-exist in invertebrate and vertebrate neurons \citep{Mulloney69, Maranto1984, Meyrand2803, Sheffield2010, Dugladze1458, Lehnert5370}. Marked differences in somatic AP recordings are observed when those are generated at the AIS or in a distal part of the axon. In particular, APs have more negative thresholds when they are generated in distal parts of the axon. This is likely due to the fact that they do not show the strong coupling of soma to AIS which is responsible for the inactivation of sodium channels at the AIS and for a more depolarized AP generation threshold. Interestingly, given the different potential at which ectopic spikes are generated, the conductances activated during AP generation may differ between distal and proximal APs \citep{Meyrand2803}.

In invertebrates, numerous examples of ectopic spikes and of their functional relevance have been described \citep{Mulloney69, Maranto1984, Meyrand2803}. Ectopic spikes in neurons that target the hearts of the leech have been proposed to control the heart firing frequency \citep{Maranto1984}. In the somatogastric ganglion of the crab, ectopic spikes are typically observed in the lateral gastric motor neuron only when the muscles remain attached to the preparation, ectopic spikes being induced by motor contraction \citep{Meyrand2803}. Remarkably, ectopic spikes in  \citep{Meyrand2803} fail to depolarize terminals onto interneurons located close to the soma, whereas they efficiently excite a distal postsynaptic target, the muscle. Thus, orthodromically-propagating spikes generated close to the soma and antidromically-propagating spikes generated distally co-exist and they can reach synapses that target different postsynaptic neurons.

Ectopic spikes have also been observed in vertebrates. In principal cells from CA3 area in the hippocampus during rhythmic activity in the gamma range \citep{Dugladze1458}, axons fire APs at five times the firing frequency detected at the soma. This is due to the activity of a specific interneuron, the axo-axonic cell, that inhibits the initial segment, thereby avoiding the backpropagation of axonal spikes to the soma. Another noteworthy case of co-ocurrence of AIS and ectopic spikes has been reported in \citep{Lehnert5370} in auditory medial superior olive (MSO) neurons. A realistic model of MSO neurons which takes into account their axonal structure and ion channel composition together with a physiological synaptic bombardment regime generated APs at both the AIS and at the first node of Ranvier. Additionally, under certain pathological conditions, e.g. in epilepsy, hyperexcitable axons have been reported to generate ectopic spikes in the hippocampus \citep{Stasheff1993}.

We would like to make at this point a clarification: although 'action potential' and 'spike' are two terms used to refer to the sodium AP generated at the axon initial segment, the term spike seems preferentially used in the literature to refer to spikes that can differ in their location (dendritic or ectopic spikes) and in their underlying ionic mechanism (e.g., calcium spike, see below). 

\subsubsection{Beyond the sodium spike}
The spike initiated through sodium influx is not the only spike propagating in the axon. There are spikes that are generated through the influx of calcium. In the giant axon of the jellyfish Aglantha digitale (order Hydromedusae), both sodium and calcium spikes propagate in the axon \citep{Mackie1985}. Sodium-dependent spikes are responsible for fast swimming and calcium spikes mediate slow swimming. Also in the axon giant neurone R2 of Aplysia, the propagating AP has a mixed Na/Ca dependency \citep{Horn1978} . 

In vertebrates, calcium spikes are typically restricted to the dendritic compartment where there is  a calcium spike generation mechanism. The biophysical mechanisms contributing to the dendritic spikes initiation in the distal apical trunk and proximal tuft of hippocampal CA1 pyramidal neurons \citep{Gasparini11046} are (1) dendritic sodium and potassium channels that set the threshold and determine the shape and forward-propagation of dendritic spikes, (2) the highly synchronized inputs (approximately ∼50 synapses activated within 3 msec) and (3) the spatial clustering (all inputs arriving in ≤ 100 $\mu$m of the dendrite). Moreover, NMDA receptors enhance spike initiation by counteracting the shunting of AMPA synaptic conductances. 
In the olfactory bulb mitral cells and in hippocampal and cortical pyramidal cells, dendritic spikes can trigger one or more axonal APs \citep{Chen1997, Stuart1997, Golding1998, Larkum1999, larkum2001, Ariav7750} . In addition, backpropagating axonal APs can themselves promote dendritic spikes, a reciprocal interaction that can lead to a burst of axonal APs \citep{Pinsky1994, Mainen1996, Larkum1999, larkum2001, Doiron2002}.

\subsection{BIOPHYSICS OF ACTION POTENTIAL PROPAGATION}

\subsubsection{An equivalent electrical circuit for axons}
The passive properties of axons can be modeled by an equivalent electrical circuit (Fig. 3A, top). The axonal membrane can be reduced to two circuit elements: the lipid bilayer, modeled by a capacitor and ion channels, by a resistor. The resistance to axial current flow can be modeled by an additional resistance.

How fast signals propagate is critically controlled by the capacitance. Electrical currents need to first charge the membrane capacitance, that opposes the flow of electric current, before electrically-charged membranes can undergo voltage changes. Capacitances are proportional to the membrane capacity (capacitance per surface area) and to the membrane surface area. The capacity of neuronal membranes is in the order of 1 $\mu$F per $cm^{2}$ \citep{GENTET2000314}. Capacitance measurements from small axons are in the range of tens of picoFahrads (pF) \citep{MEJIAGERVACIO2007843}. Capacitances of larger neurons (i.e., strongly-ramifying interneurons and projection neurons) are therefore expected to be in the range of hundreds of pF to the nF range. Dynamic changes in capacitance are suggested by activity-dependent changes in the size of axons (e.g.,\citep{chereau2017superresolution}). Little is known about dynamical changes in capacitance due to temperature, lipid composition, and whether these may significantly impact the propagation of electrical signals along the axon.

The membrane resistance plays a major role in controlling how far signals spread in space along the axon before membrane potential changes become imperceptible. This is typically measured by the space constant, defined as the distance over which the membrane potential decays to 37 percent of its initial value. The more open channels are available at the membrane, the more the axial current is attenuated in space along the axon due to current leakage through membrane channels. Remarkably, the membrane resistance can change (reviewed in \citep{DEBANNE201973}), providing a potential source of variation in the conduction of electrical signals along axons. 

The axonal axial resistance to current flow results from the combination of the resistivity of the axoplasm (that is, the cytoplasm of the axon), given in Ohms*cm, and the diameter of the cylinder. Axial resistivity values are in the range of 〜 100 $\Omega$.cm \citep{carpenter1975resistivity, cole1975resistivity}. Computational models of the calyx of Held suggest that the latency and amplitude of signals propagating between release sites is highly sensitive to changes in axial resistivity, and that these changes may have a large impact on synaptic release \citep{SPIROU2008171}. However, changes in resistivity have not been reported so far experimentally. The axon diameter can vary in three orders of magnitude, from around hundred nanometers to hundreds of micrometers in diameter. Giant fibers found in invertebrates are an example of specialized large-diameter structures that propagate electrical signals with high speed over long distances ( in \citep{Hodgkin1952,Xu1979}, Fig. 4A). The larger the diameter of the axon, the faster the electrical signal propagates. Taking into account cable theoretic considerations, the  propagation speed increases with the square root of the diameter. Remarkably, the axon diameter, as well as the diameter of synaptic boutons, are not static properties of axons. In fact, axon diameter has been shown to be regulated in hippocampal principal cells in an activity-dependent manner. Plasticity protocols induced changes in axonal diameter which were accompanied by significant changes in AP conduction velocity along CA3 pyramidal cell axons \citep{chereau2017superresolution}(Fig. 5A). 

\subsubsection{Propagation of subthreshold signals}
Subthreshold signals propagate passively in axons, reaching synaptic terminals, where they influence synaptic release \citep{alle2006combined}. Subthreshold membrane fluctuations consist of synaptic events, typically in the range of a 100s of $\mu$Vs to mVs, which in cortical neurons approximate highly stochastic background dynamics in vivo \citep{rudolph2003characterization}. Their propagation, referred to as analog, in comparison to the digital nature of the AP, modulates the efficiency of APs to induce neurotransmitter release in synaptic terminals \citep{Shu2006}. Given that subthreshold signals are not regenerated along the axon, analog signaling is more prominent in proximal portions of the axon. The presence of combined analog and digital signals in axons has been termed hybrid analog-digital signaling.

Of particular interest is that the propagation of slow analog subthreshold signals does not only reach chemical synapses. Subthreshold signals also reach axonal electrical synapses, at which the signal is expected to be conveyed with high efficiency to the postsynaptic site due to the continuous and low-pass filtering properties of electrical transmission (reviewed in \citep{alcami2019beyond}).

\subsubsection{Reliability of propagation and action potential failures}

Before we consider how signals propagate along axons, let us discuss the reliability of propagation. APs do not always efficiently propagate, and in fact, failures in AP propagation are observed in all neuron types at high firing frequencies \citep{Krnjevic1959, Grossman1979, Monsivais464}.

Although geometry can generate failures (e.g. at branch points, see next section), failures typically involve active mechanisms during repetitive activation of axons as observed in a number of vertebrate and invertebrate preparations \citep{Krnjevic1959,Monsivais464,Mar4335}. Repetitive activity typically leads to extracellular potassium accumulation and subsequently to the depolarization of the axon, inducing failures of conduction \citep{Grossman1979}. In other cases however, repetitive activation induces AP propagation failures via hyperpolarization of the membrane  \citep{Mar4335}. Note that propagation failures likely affect only a fraction of APs at physiological firing rates, and that most APs succeed in propagating.

Interestingly, changes in AP failure can also occur in response to membrane fluctuations, e.g. in response to synaptic inputs. Indeed, specific potassium channels of the type $I_{A}$ underlie a hyperpolarization-mediated conduction block in hippocampal pyramidal cells \citep{debanne1997action}. Remarkably, the activation and de-inactivation kinetics of $I_{A}$ allow for a history-dependent conduction block of APs.

\subsubsection{More than linear cables: Impact of axonal branching and inhomogeneities}
Axons are typically formed by complex trees and spatial heterogeneities. These include branching points, varicosities (local enlargements of the axon containing the release machinery of chemical presynaptic sites) and large structures specialized in the interaction with other cells. Examples of such structures are the ‘basket’ formed by a specific type of interneuron, basket cells, around principal cells in many brain regions; the glomerular collateral of the climbing fiber in the cerebellum; the ‘pinceau’ structure surrounding cerebellar Purkinje cells \citep{palay2012cerebellar} ; Fig. 1B, C) or the calyx of Held in the auditory system.

Changes in axon diameter at varicosities or branching points are characterized by an impedance mismatch, that is, a need of a larger current to flow in one of the two directions to electrically load axonal branches, provoking changes in AP propagation \citep{GOLDSTEIN1974731, MANOR19911424}. Axonal branches indeed represent a local challenge to the propagation of electrical signals, due to the impedance mismatch between the mother and daughter branches. That is, if APs need to load a larger impedance as they propagate from one branch into two branches, their propagation will be delayed relative to the speed that they would have had if no branching was present.

In the most extreme case, the electrical signal fails to load one or two of the branches, resulting in AP propagation failure. Propagation failures have been shown to occur in branches of a number of neurons \citep{Yau1976, Gu1991, Grossman1979}. An interesting example is provided by the medial pressure sensory neuron in the leech, where the failure of APs to propagate can differentially affect postsynaptic cells contacted by distinct presynaptic branches \citep{Gu1991}.

Finally, branching points can also provoke a surprising effect: APs can be slowed down in the ms range, up to a level that allows the mother branch to overcome the refractory period for AP generation. As a consequence, the AP can 'reflect' or travel backwards, increasing synaptic release at synapses present in the branch where the AP reflects \citep{baccus1998synaptic, baccus2000action}.

The impact of complex morphologies on AP propagation is likely to be relevant in the complex axonal ramification patterns of many neurons, including vertebrate interneurons \citep{Ofer414615}. The usage of voltage sensitive-dyes that track membrane voltage with sub-millisecond precision \citep{Palmer1854} should allow following large portions of axons both in vitro and in vivo, and characterize the propagation of APs along complex axonal structures.

\subsubsection{Biophysical properties of myelinated fibers}
Many vertebrate and some invertebrate fibers are myelinated, a specialization endowed by the glial ensheathment of axons \citep{castelfranco2015evolution}. Myelin, formed by compact lipidic layers produced by the membranes of glial cells (oligodendrocytes in the central nervous system and Schwann cells in the peripheral nervous system), strongly impacts the propagation of electrical signals.

In his seminal study, Lillie \citep{Lillie1925} wrapped an iron wire placed in an acidic solution with an insulating glass cylinder. This increased the speed of propagation of electrical signals along the wire, which he postulated to occur in myelinated fibers. His prediction was confirmed decades later in axons by recording electrical signals at nodes, leading to the concept of the ‘saltatory’ nature of transmission of electrical signals in myelinated fibers \citep{Huxley1949}. Saltatory comes from the latin verb 'saltare' (to jump), an analogy describing the very fast propagation of electrical signals between nodes of Ranvier.

Myelin modifies the electrical circuit that models the passive properties of axons described above. It increases the effective radial resistance and decreases the effective capacitance of the axon (Fig. 4). This effect is due to the different properties of series resistors and series capacitors added to the circuit: series resistors sum their resistances and the inverse of capacitances from capacitors in series sums. The increase in effective membrane resistance \citep{bakiri2011morphological} and the decrease in effective membrane capacitance by myelin have two major consequences. On the one hand, the increase in the effective axonal resistance by myelin increases their length constant. On the other hand, as a consequence of the decrease in the effective axonal capacitance, the time required to effectively load axons decreases, dramatically accelerating the propagation of electrical signals. This speeding of electrical propagation by a reduced effective capacitance underlies saltatory conduction between nodes of Ranvier \citep{Huxley1949, castelfranco2015evolution} .

\subsubsection{Geometry of axons and myelin}
Specific myelination parameters have been shown to maximize conduction velocity of electrical signals along axons. In particular, the ratio of the axonal diameter d to the fiber diameter D (the summed diameter of axon and myelin sheath), defined as the 'g-ratio', has been shown to critically control the conduction of electrical signals. Rushton \citep{Rushton1951} developed a biophysical formalism to model current flow in a myelinated axon. He deduced the relation between the ratio l/D (internode length l over D) and the g-ratio. He further demonstrated that the ratio l/D is maximal when g = 0.6, a value also found analytically to maximize the space constant. In an independent approach, Deutsch \citep{deutsch1969} mathematically derived the geometry of axonal and myelin properties that maximize, this time, conduction velocity. He deduced that the propagation velocity is inversely proportional to the internode length and to the RC time constant given by the internal resistance of the axon and the capacitance of the membrane. Maximizing conduction speed requires minimizing the time constant of the circuit. This resulted in the same geometrical properties as those derived by Rushton: a g-ratio of 0.6. Additional modeling studies including a more complete description of myelinated fibers converged to similar conclusions \citep{GOLDMAN1968596}. Therefore, due to the biophysical constraints posed by the thickness of myelin and axonal size, both conduction speed and efficient spatial propagation of electrical signals are maximized by specific axonal and myelin geometries. It is noteworthy that the geometry of axons and myelin does not only control AP propagation speed \citep{Rushton1951, deutsch1969} but also their temporal jitter (Kim et al., 2013).

It seems difficult to imagine that neurons have fine-tuned their dendritic computations in a cell-type specific manner \citep{stuart2016dendrites}, but that axons would, on the contrary, be invariant and homogeneously optimizing speed by their geometry. Additionally, we know that nervous systems adapt and fine tune a plethora of properties to accomplish specific functions \citep{Marder2011}. Axons indeed adjust different parameters which impact the speed of propagation of electrical signals \citep{Seidl70, ford2015, arancibia2017node} to obtain different speeds of computation.

An illustrative example of how the geometry of axons and myelin is tuned to adjust AP propagation speed, deviating from a g-ratio of 0.6, is provided by the axon properties which encode spatial location in the avian brain \citep{Seidl70}. The longer contralateral fibers have larger-diameter axons and longer internodal distances, compensating in this manner for an otherwise slower conduction delay relative to the shorter fibers on the ipsilateral side. In this manner, axons ensure a coincident arrival of contralateral and ipsilateral signals to the synaptic terminals. Another interesting example is provided by fibers specialized in carrying information for low-frequency sounds, which show larger diameters than those carrying high-frequency sounds, but also shorter internodes. In doing so, these fibers deviate from the classical dependence of both variables established by Rushton. This specialization is proposed to ensure a proper function of the circuit \citep{ford2015}. Moreover, internode length and node diameter show graded properties as they approach the terminal in the auditory granular bushy cell axon, implementing an efficient invasion of the axon terminal by APs \citep{ford2015}.

We have seen that a large number of structural myelination parameters act in concert to control the speed of AP propagation \citep{GOLDMAN1968596}. It is noteworthy to mention that myelin has introduced new degrees of freedom in the regulation of AP speed: speed depends on distance between nodes of Ranvier, on node length, node composition, and thickness of myelin \citep{Rushton1951, ford2015, arancibia2017node, wu2012increasing}. These parameters can be modulated independently or in combination, thereby increasing the number of available mechanisms by which nervous systems tune the axonal propagation of electrical signals. Furthermore, dynamical changes in axons and myelin \citep{Sampaio-Baptista19499, McKenzie318, Sinclair8239, field2015} suggest that nervous systems dynamically adjust the properties of their fibers to achieve the specific behaviors that they control. 

\subsubsection{Active contributions to variations of action potentials along the axon at nodes and terminals}
Myelinated fibers are characterized by the presence of non-myelinated sections, which concentrate the machinery to generate APs. These include nodes, the heminode (last unmyelinated portion before the terminal) and synaptic terminals. Remarkably, AP shape can suffer modifications along the axon by being modulated at each one of these loci by the active channels that they express. Each active AP regeneration site along the axon can potentially generate AP variants due to their specific state and composition of ion channels and membrane potential.

An example of local variations in AP shape through active mechanisms is found at presynaptic mossy fiber terminals (present in the axon of dentate gyrus granule cells), where local potassium channels modulate spike shape \citep{alle2011sparse}. Another example is the activation of calcium-dependent potassium channels in Purkinje cell nodes of Ranvier. These channels repolarize membranes, de-inactivating sodium channels which can then generate fast frequency spikes, and in this manner prevent AP failure at these high frequencies \citep{GRUNDEMANN20151715}. Additionally, changes in membrane voltage in the axon induced by active mechanisms can impact the efficiency of AP propagation. For example, in the leech touch cells, adaptation in response to repetitive AP firing consists of a hyperpolarization, resulting in the blockade of AP propagation \citep{Van1973}.

We previously described how actively-generated signals can differently be passed on by two bifurcating branches, illustrating how passive properties can filter actively-generated signals \citep{Gu1991}. Remarkably, active properties of axonal branches can prevent failures by potentiating signals in specific branches \citep{cho2017sodium}. Failures that would occur as a consequence of the passive filtering of electrical signals can be prevented in a branch-specific manner by the presence of the sodium channel subunit Nav$\beta$II, which potentiates AP propagation.

\subsubsection{The geometries of axons and myelin are plastic}
Geometrical properties of both axons and myelin have been shown to be highly plastic. Indeed, the diameter of axons can change as a function of neuronal activity \citep{Sinclair8239, chereau2017superresolution}. The understanding of the mechanisms of myelination has revealed that properties of nodes and internodes are subject to plasticity \citep{Young2013}, reviewed in \citep{KALLER201786}. Interestingly, major macroscopic changes in myelination occur in response to training paradigms in mice, allowing the acquisition of motor skills \citep{Sampaio-Baptista19499, McKenzie318, Xiao2016}, and models of injury induce strong remodelling of myelination patterns in the auditory brainstem \citep{Sinclair8239}. Changes in myelination can involve several mechanisms at time scales of hours to days: changes in internode length \citep{Etxeberria6937}, in myelin thickness \citep{Sinclair8239}  and in node length (suggested in \citep{arancibia2017node}, (Fig.5).

\subsubsection{Nanostructures with unknown function}

The evolutionary appearance of myelin has led to new specialized microstructures or microdomains (namely nodes, paranodes, juxtaparanodes, heminodes). Electron microscopy and recently superresolution imaging have revealed additional structures whose contribution to axonal computation remains mysterious. These are organized at the nanometer scale in a highly regular spatial organization \citep{D'este2017}, rising the question of the function of these periodic structures. For example, d'Este et al. show that the voltage-dependent potassium channel subunit Kv1.2 channels, found at the juxtaparanodes, correlates in space with the underlying actin cytoskeleton. An additional structure that has attracted our attention is the Schmidt-Lanterman incisure, a spiral cytoplasmic expansion from the outer tongue of myelin to the inner tongue. Incisures express gap junctions and their contribution to the electrical properties of myelin remain mysterious \citep{KAMASAWA200565}. Likewise, exquisite arrangements of structures apposing glial and neuronal membranes and also with other glial membranes in the form of ‘rosettes’ formed by ion channels at the paranode are still poorly understood \citep{Rash2016}. 

These findings open up the following question: are these highly-ordered structures an epiphenomenon resulting from the necessary anchoring to regular cytoskeletal structures, or is this spatial regularity impacting function? One would expect that further perturbing and studying those nanoscale structures in the future will further our understanding of how they contribute to axonal computations.

\subsection{AXONS ARE NOT ALONE: INTRACELLULAR AND INTERCELLULAR COUPLING}

\subsubsection{Crosstalk of axons with other compartments}
Although we have done a treatment of the axon as an isolated cable as has classically been performed in the early days \citep{Lillie1925, Huxley1949, Hodgkin1952}, the axonal cable is coupled to the somatic and dendritic compartment at one end and, additionally, directly or indirectly to electrically-coupled cells, which also functionally act like an electrical compartment \citep{Furshpan1959, alcami2013estimating, Eyal8063}. The electrical coupling between the axon and the soma, and indirectly to dendrites and dendritically-coupled cells, influences AP generation in the axon \citep{bekkers2007targeted, Eyal8063, amsalem2016neuron, alcami2018electrical, goldwyn2019soma}. These compartments create parallel pathways for current flow that compete with the flow of currrent in the axial resistance of the axon. Loading of these non-axonal compartments was shown to impact AP threshold and speed \citep{bekkers2007targeted, Eyal8063, amsalem2016neuron}. Their contribution to the effective membrane time constant is substantial. The somato-dendritic compartment has been shown to, by this mechanism, modify both the threshold and the initial rise of the AP \citep{Eyal8063} .

It is further interesting to consider the axon in the context of synaptic integration, that is, the computation of information received at synapses. Similar to the somato-dendritic and junctional compartments, which act as current sinks, influencing the effective kinetics and strength of excitatory inputs recorded at the soma, before they reach the AIS \citep{Norenberg2010, alcami2018electrical}, the axonal membrane also behaves as a current sink, leaking current generated at synapses in the somatic and dendritic compartments with its charging time constant. This phenomenon has been shown to, through a passive mechanism, accelerate the time-course of somatically-recorded excitatory synaptic events \citep{MEJIAGERVACIO2007843}. It has additionally been suggested to also accelerate the time-course of spikelets generated by electrically-coupled cells \citep{alcami2013estimating}. Mejia-Gervacio and collaborators \citep{MEJIAGERVACIO2007843} show that the capacitive loading of axons introduces in cerebellar molecular layer interneurons a computational time constant of about 3 ms, which is one order of magnitude slower than the faster time constant to load the somatodendritic compartment. This relatively slowly-charging process of axons accelerates the decay of excitatory postsynaptic potentials, reducing the time window for AP generation. Therefore, electrical signals that arrive to the axon are not only passively influenced by other compartments, but they also influence passive computation by the non-axonal compartments, as part of a system formed by coupled compartments.

Let us now turn our attention onto the signals propagating in axons as cells detect events in their somas and dendrites, which is a consequence of the current flow of synaptic events into the axon. These signals propagate at long distances before their complete attenuation (contributing to the analog signaling in the axon that was previously introduced). The space constant at the hippocampal unmyelinated granule cell axon is in the range of hundreds of micrometers \citep{alle2006combined} and the subthreshold propagation of voltage depolarizations has been shown to increase AP evoked release \citep{Shu2006}. Interestingly, these signals do not only travel orthodromically towards the terminals, but signals generated in the axon, e.g. by synaptic receptors present on the presynaptic membrane, also travel antidromically to the soma \citep{TRIGO2010235}. As a consequence, these axonal PSPs depolarize the soma and influence AP generation (de San Martin et al., 2015).

\subsubsection{Direct coupling between axons}
Direct coupling between axons was suggested in early work, bringing up the concept that networks of axons may directly interact with each other (Katz and Schmitt, 1940). These interactions can be explained by two forms of electrical transmission (Fig. 6A): ephaptic transmission due to the generation of an electric field by an axon, affecting the excitability of a neighboring axon \citep{Katz1940, blot2014ultra, HAN2018564}, and transmission mediated by gap junction-mediated electrical synapses \citep{SCHMITZ2001831, Furshpan1959, Watanabe267, bennett1963electrotonic, Robertson1963}. As a matter of fact, electrical synapses were first discovered in axons in a variety of preparations \citep{Furshpan1959, Watanabe267, bennett1963electrotonic, Robertson1963} .

Axonal electrical synapses are expected to strongly affect electrical signals by adding a conductance pathway directly in the axon. Remarkably, axonal gap junctions allow inputs to arrive at the axon directly, blurring the pure ‘output’ role of axons: an AP in a presynaptic axon induces a spikelet (a low-pass filtered version of the presynaptic AP) in the postsynaptic axon. Spikelets excite distal parts of the axon where the gap junctions are located, and when the spikelet depolarizes the membrane potential sufficiently, they are able to evoke APs \citep{chorev2012vivo}. Interestingly, in vivo recordings from hippocampal principal cells in CA regions from rodents revealed two types of APs: one type was preceded by a ‘shoulder’ which was identical to the rising phase of the spikelet, and the other was a full-blown AP lacking the shoulder \citep{Epsztein474}. Spikelets correlate with AP firing from nearby cells, confirming that they are generated by electrically-coupled cells, and not by spontaneous AP firing of distal axons \citep{chorev2012vivo}. Altogether, these studies suggest that hippocampal pyramidal cell axons are excited through electrical synapses in physiological conditions in behaving animals, as it had been previously shown in slices \citep{SCHMITZ2001831} .

Axonal gap junctions have additionally been proposed to underlie fast synchronization of neuronal ensembles in both physiological and pathophysiological conditions \citep{SCHMITZ2001831, Roopun338}. Interestingly, axo-axonal gap junctions, by coupling axons, that is, two output structures, can have a strong impact on spike coordination and coding efficiency \citep{Wang2017} .

Last and not least, direct axo-axonal coupling can be mediated by chemical synapses onto axons (Fig. 6A) which are typically performed by a specific type of interneuron targeting the AIS of principal cells (Somogyi et al., 1998). It has been shown that the presynaptic activity of chandelier cells modulates the sodium channel dynamics on the axon initial segment \citep{inda2006voltage}. Terminals found in the neocortex and hippocampus were found to exert inhibitory control over AP initiation \citep{Dugladze1458}.

\subsubsection{Axons interact indirectly via glial cells}
Finally, let us consider the contribution of glial cells to axonal function as an indirect coupling pathway between axons (Fig. 6B). The biophysical properties of myelin are achieved by glial cells that produce myelin. Since several axons are typically contacted by a myelinating cell, this introduces de facto an indirect coupling pathway between axons. A number of transmission mechanisms has been described between axons and myelin (reviewed in (\citep{Micu2017})), whose dynamic properties depend on signaling from neurons \citep{Hines2015}.

\section{Conclusion}

Here, we have reviewed the biophysical nature of computations performed by the axon. We have shown that the axon is not just a cable on which electrical pulses propagate but rather a computational device that modulates signaling and adds to the complexity of information processing in the brain. It should be appreciated that these computations are sophisticated enough to contribute to network level phenomena up to behavior in vivo. Relating axonal computation to behavioral phenomenology is still a nascent area but it will complement studies that have focused almost exclusively on somas, dendrites or a coarse grained view of the neurons. The study of axonal computation is hurdled by technical challenges, but there is already an emerging interest in developing technologies that will allow to electrophysiologically and structurally study axonal processes. Further complication arises when one realizes that axons do not act alone but in concert, exhibiting collective modes of computation conveyed by inter-axonal signaling. We propose that axonal biophysics are of vital importance for understanding not only how single neurons process information but also how neural networks coordinate their activity.

\section*{Conflict of Interest Statement}

The authors declare that the research was conducted in the absence of any commercial or financial relationships that could be construed as a potential conflict of interest.

\section*{Author Contributions}

Both authors contributed equally to literature search and writing.

\section*{Funding}
PA is funded by the Munich Center for Neurosciences. AEH is funded by the Howard Hughes Medical Institute. 

\section*{Acknowledgments}
We are grateful to Conny Kopp-Scheinflug for her feedback on previous versions of this manuscript.

{ \bibliography{axon.bib}}
\bibliographystyle{apalike}
\newpage
\section*{Figures}


\begin{figure}[h!]
\begin{center}
\includegraphics[width=10cm]{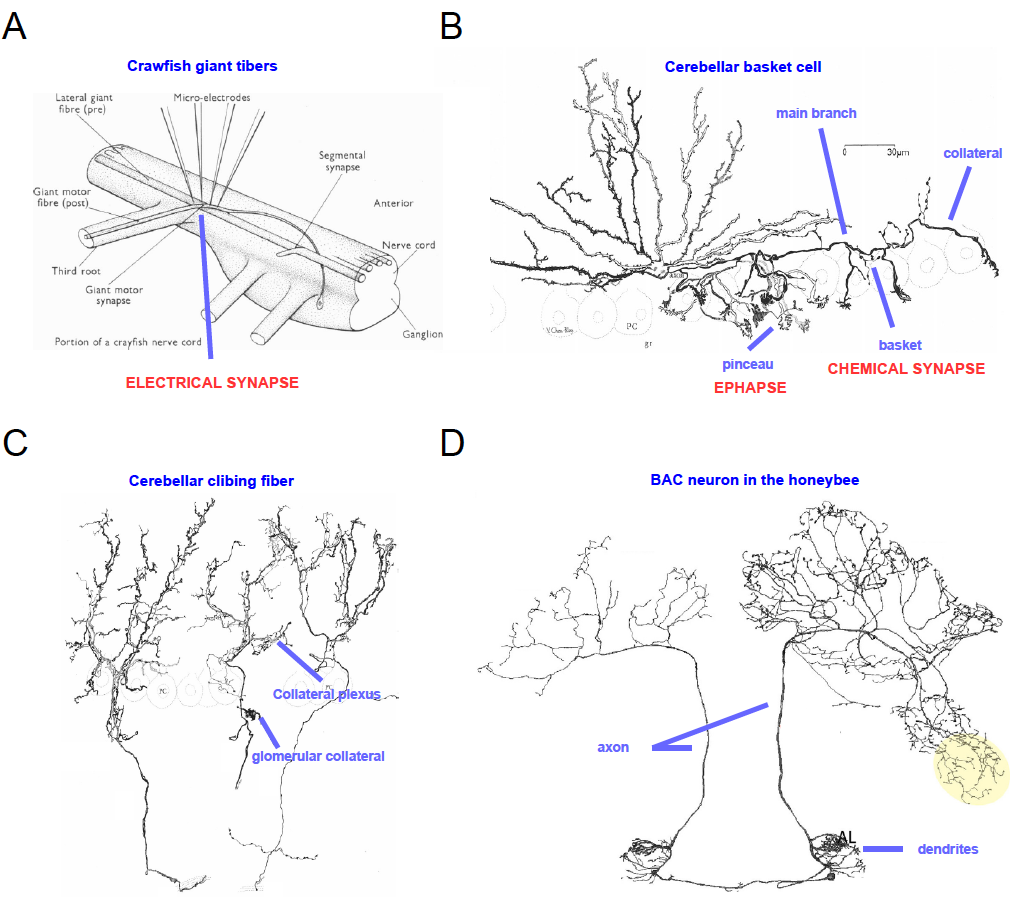}
\end{center}
\caption{ Diversity of axons: morphology and function. A. Giant fibers in the crawfish contact each other
by electrical synapses. From Furshpan and Potter, 1958. B. Cerebellar basket cells collaterals contact
several Purkinje cells, which they inhibit at the basket structure via chemical synapses, and the ’pinceau’
via ephaptic inhibition. From Palay and Chan-Palay, Springer Verlag Berlin, 1974. C. The climbing fiber in
the cerebellum shows different axonal specializations: the glomerular collateral in the granule cell layer
and the climbing fiber wrapping around Purkinje cells. From Palay and Chan-Palay, Springer Verlag Berlin,
1974. D. The honeybee axon of the BC cell projects onto the left and right hemispheres, targeting a number
of regions (one of the target regions is labelled in yellow). From Zwaka et al., 2016.}\label{fig:1}
\end{figure}

\begin{figure}[h!]
\begin{center}
\includegraphics[width=10cm]{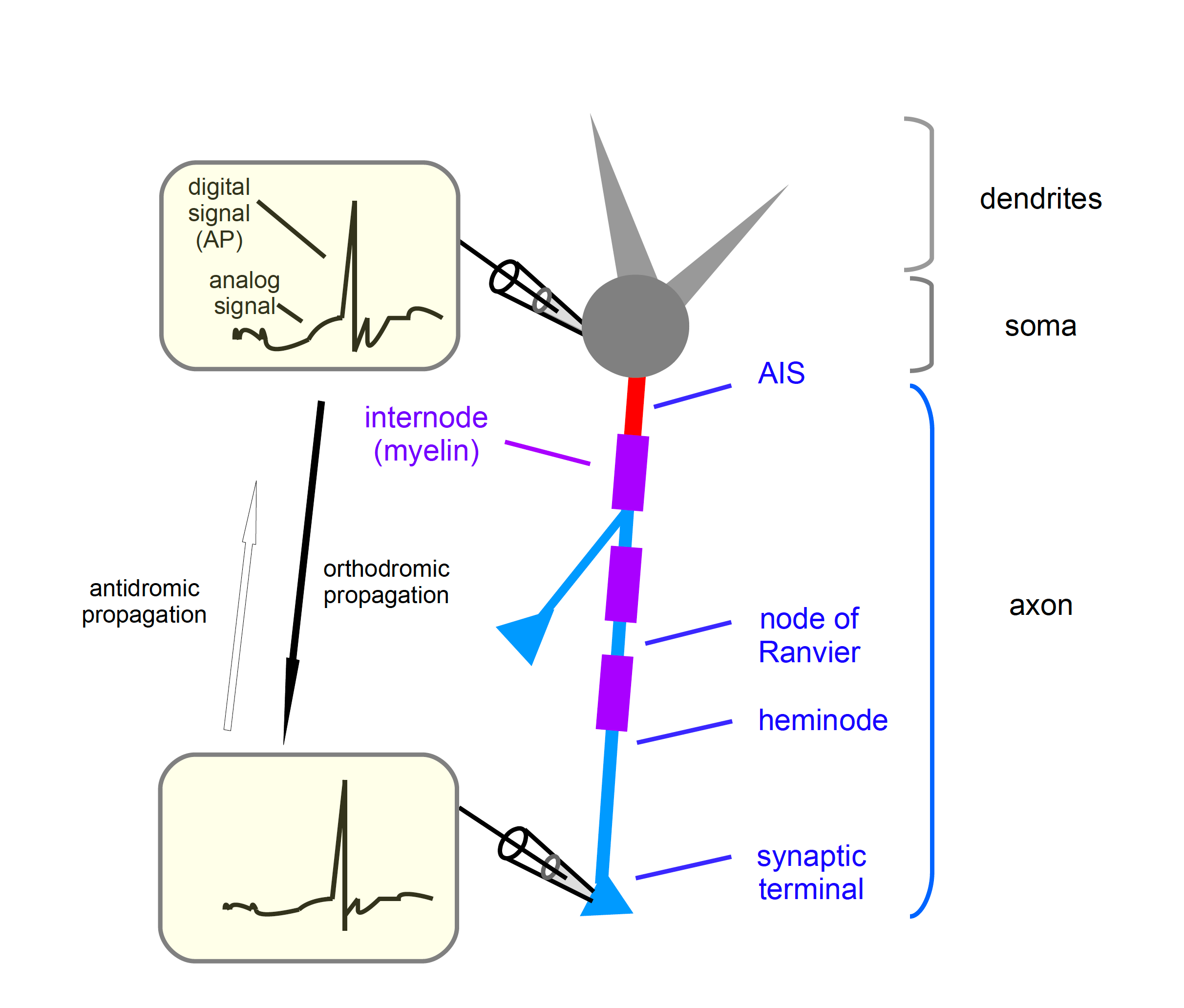}
\end{center}
\caption{ Overview of axonal electrical signaling. Axonal signaling is formed by a mixture of analog and
digital signaling. Insets: membrane potential traces. Action potentials can propagated orthodromically or antidromically. Right, in blue, active sites with the machinery to generate or regenerate action potentials. AIS, axon initial segment.}\label{fig:2}
\end{figure}

\begin{figure}[h!]
\begin{center}
\includegraphics[width=20cm]{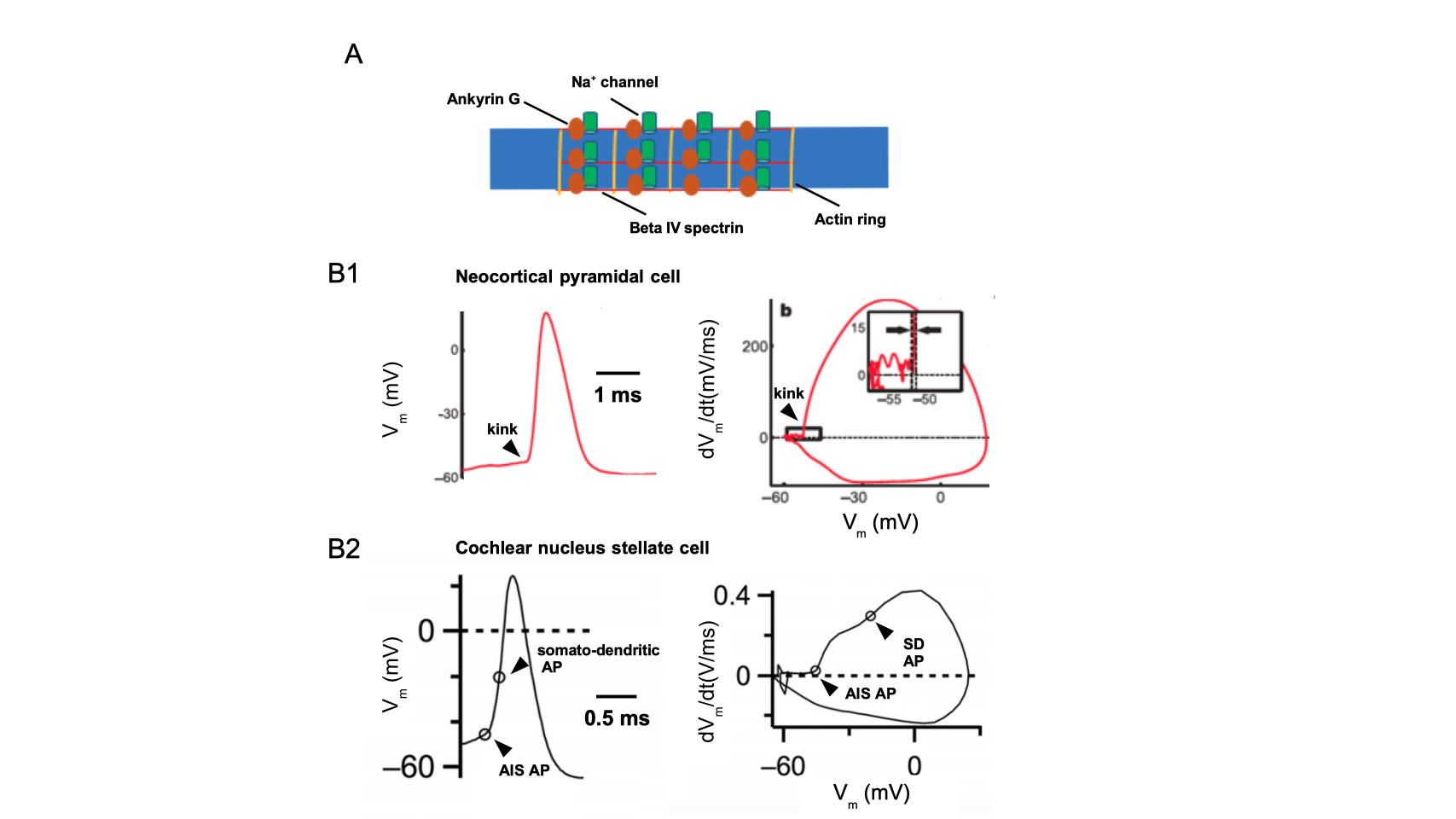}
\end{center}
\caption{ Overview of Action Potential Diversity. A Ultrastructure of the axon initial segment. Green is sodium channels, orange circles are Ankryin G molecules, red lines are beta IV spectrin and orange lines are the actin rings. B1. Left, example of fast rising action potential from a neocortical pyramidal cell. Right, phase plane plot of this cell. B2. Left, example of a stellate cell from the cochlear nucleus with two components: a fast rising phase of the action potential contributed by the AIS and a second phase contributed by sodium by the somatodendritic compartment. Left, phase plane plot of this cell. }\label{fig:2}
\end{figure}

\begin{figure}[h!]
\begin{center}
\includegraphics[width=10cm]{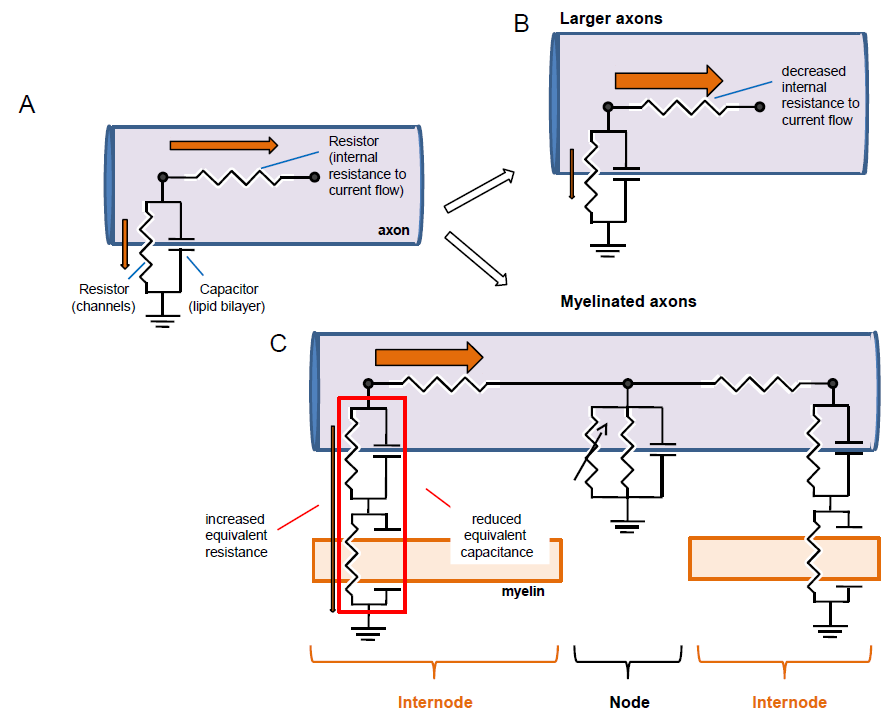}
\end{center}
\caption{ Equivalent circuits of axons and structural biophysical specializations that improve conduction
Top left. An axon can be reduced to an equivalent electrical circuit involving an axial resistor (parallel to
the membrane) conveyed by the intracellular medium of the axon (axoplasm), and a parallel ‘RC circuit’
formed by the membrane resistance and membrane capacitance. Top right, conduction can be improved
by decreasing axon diameter, and thereby longitudinal resistance to current flow, allowing for a larger
current flow in the axon relative to the membrane resistor, and thereby a larger space constant and faster
conduction. Bottom, an alternative circuit modification occurs with myelination: additional capacitances conveyed by myelin
in series with axonal capacitance reduce the effective capacitance and additional resistance increase the effective resistance,
increasing propagation speed and space constant, respectively. Note that the battery associated to each
membrane resistor has been omitted for simplification purposes. Orange arrows represent the current flow,
and their thickness is indicative of the relative current flow in the axoplasm and in the radial direction out
of the axon.}\label{fig:4}
\end{figure}

\begin{figure}[h!]
\begin{center}
\includegraphics[width=10cm]{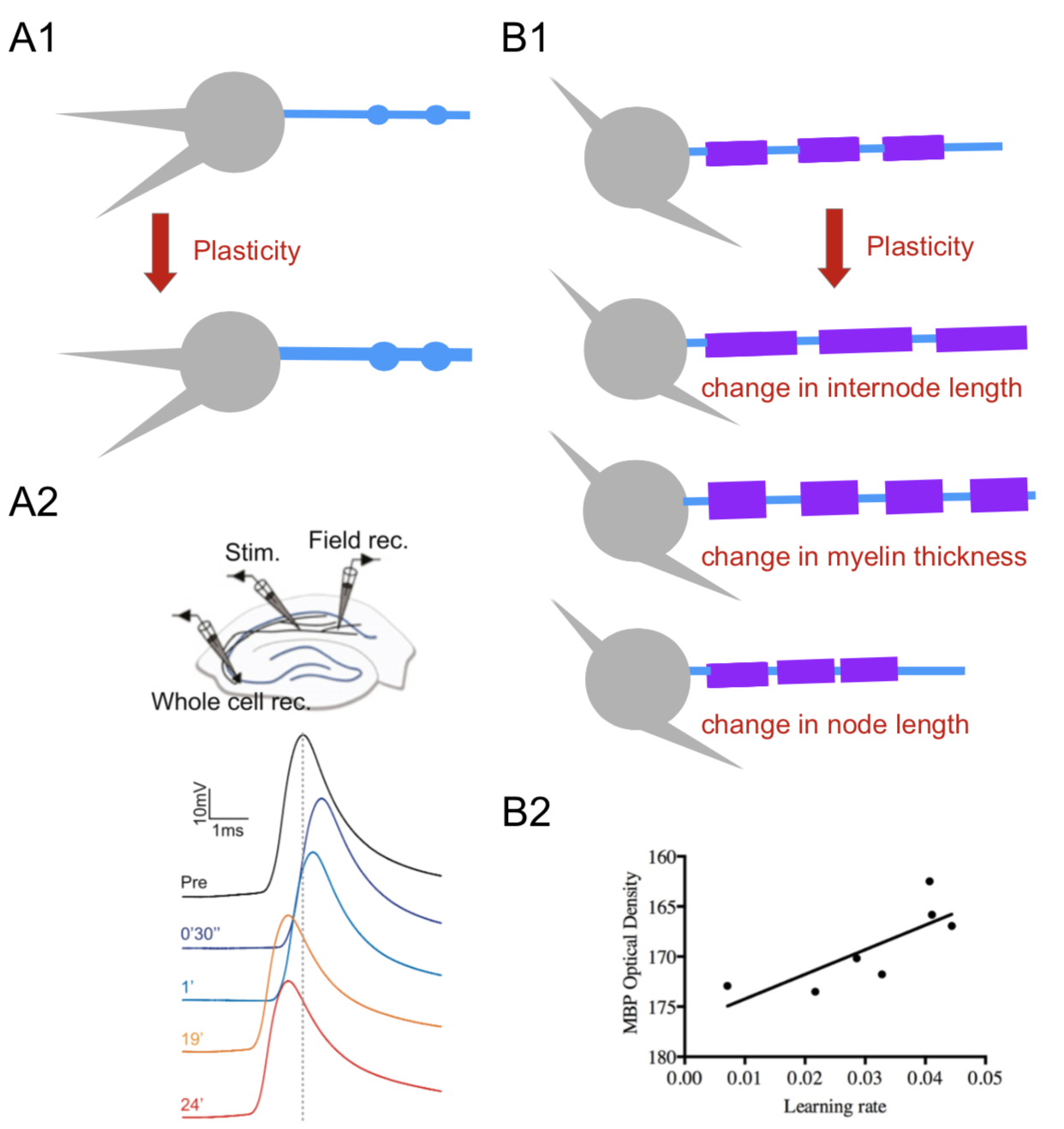}
\end{center}
\caption{ Structure-function plasticity A. A1. Axon structural plasticity. A2. Functional implications:
action potential latency, measured by antidromically stimulating axons before and after a plasticity protocol,
decreases several minutes after the protocol. From Ch´ereau et al. (2015) B. B1. Myelin structural plasticity
can involve changes in internode length (Etxebarria et al., 2016), in myelin thickness (Sinclair et al., 2017)
or in node length (suggested in Arancibia-Carcamo, 2017). B2. Functional implications of changes in myelination: MBP
optical density, putatively proportional to the amount of myelin positively correlates in the learning rate in
a motor skill learning paradigm. From Sampaio-Baptista et al. (2013).}\label{fig:1}
\end{figure}

\begin{figure}[h!]
\begin{center}
\includegraphics[width=10cm]{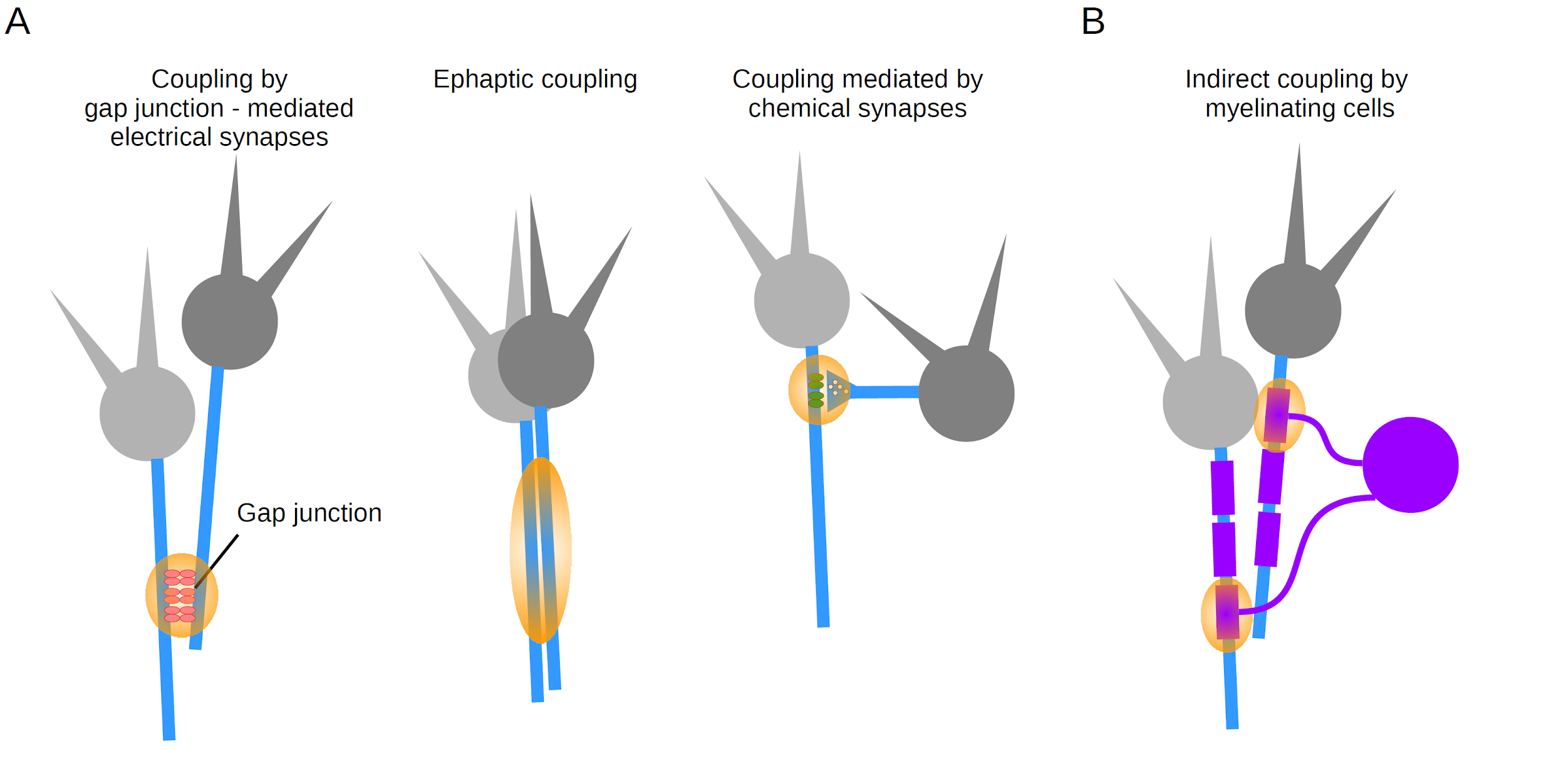}
\end{center}
\caption{Coupling modalities between axons. A. Three types of synapse-mediated coupling: through axo-axonal gap junctions (left), ephaptic coupling (middle), and chemical axonic synapses (right). B. Indirect coupling via myelinating cells.}\label{fig:1}
\end{figure}


\end{document}